\documentclass[osajnl,twocolumn,showpacs,10pt]{revtex4-1}

\usepackage{amsmath,amssymb,graphicx}
\usepackage{dcolumn}

\begin{document}

\title{Thermal conductivity reduction by acoustic Mie resonance in nanoparticles}

\author{Brian Slovick}\email{Corresponding author: brian.slovick@sri.com}
\author{Srini Krishnamurthy}
\affiliation{Applied Sciences Division, SRI International, 333 Ravenswood Avenue, Menlo Park, CA 94025, USA}

\date{\today}

\begin{abstract}
We evaluate the impact of acoustic Mie resonance in nanoparticles on the thermal conductivity of semiconductor and polymer composites. By appropriately choosing the bulk modulus and density, and selecting the size of the nanoparticle to align the Mie resonances with the dominant portion of the thermal conductivity spectrum, we show that large reductions in thermal conductivity are achievable with dilute concentrations of nanoparticles. In semiconductor alloys, where the spectral thermal conductivity is known, our model can explain the steep reductions in thermal conductivity observed previously. However, the results of our effort to evaluate acoustic Mie resonance in polymer composites are inconclusive due to uncertainties in the spectral thermal conductivity. Acoustic Mie resonances can be useful for maximizing $ZT$ for thermoelectric applications, since a dilute loading of nanoparticles can reduce thermal conductivity with minimal impact on electrical conductivity.
\end{abstract}

\maketitle

Materials with low thermal conductivity are useful for many applications \cite{Hu2007,Maldovan2013}. Transparent materials with low thermal conductivity are desired for window insulation \cite{Shimizu2016,Liu2018}, while those with low thermal conductivity and high electrical conductivity are useful for thermoelectric (TE) applications \cite{Slovick2016}. Efficient TE devices are desired for applications such as energy harvesting of waste heat and solid-state refrigeration \cite{Chen2003,Snyder2008}. For optimum efficiency, TE devices require materials with large $ZT$, given by $\sigma S^2 T/\kappa$, where $\sigma$ is the electrical conductivity, $S$ is the Seebeck coefficient, $T$ is the temperature, and $\kappa$ is the thermal conductivity. Approaches for increasing $ZT$ include increasing $\sigma S^2$ by doping \cite{Snyder2008} or decreasing the lattice contribution of $\kappa$ by nanostructuring \cite{Venkatasubramanian2001,Yu2010}.

Methods of reducing $\kappa$ by nanostructuring include superlattices \cite{Simkin2000,Venkatasubramanian2001} and nanoparticles \cite{Zide2005,Kim2006a,Mingo2009}. Superlattices are periodic materials comprised of layers with different acoustic properties. It has been suggested that coherent Bragg scattering between layers of the superlattice can be used to obtain a phononic bandgap, preventing phonons of a selected range of frequencies from being transmitted through the material \cite{Simkin2000,Venkatasubramanian2000}. However, practical reports indicate that $\kappa$ in superlattices is dominated by diffuse interface scattering rather than coherent Bragg interference \cite{Venkatasubramanian2001,Yang2003,Yu2010}.

An alternative approach for lowering thermal conductivity is to randomly distribute nanoparticles into the lattice \cite{Zide2005,Kim2006a,Mingo2009}. In contrast with Bragg scattering in superlattices, scattering by randomly-distributed nanoparticles is uncorrelated and relatively insensitive to disorder \cite{Kim2006a}. Of particular interest is the demonstrated 50\% reduction of $\kappa$ for InGaAs alloys containing only 0.3\% randomly dispersed semimetallic ErAs nanoparticles \cite{Kim2006a}. Owing to the small volume fraction of ErAs, the electrical conductivity of the composite was largely unaffected, leading to similar improvement in $ZT$. The results were explained using a parametrized Rayleigh approximation for the acoustic scattering cross section of a sphere \cite{Kim2006a,Kim2006b,Mingo2009}. However, such a sharp reduction of thermal conductivity for a dilute loading strongly suggests the presence of an additional scattering mechanism. Here we argue that acoustic Mie resonance may be responsible for the observed reduction in $\kappa$. 

In electromagnetics, Mie scattering is well known and has been applied to obtain near-perfect specular reflection with dense metamaterial layers \cite{Slovick2013,Moitra2014,Moitra2015} and high diffuse reflectivity with dilute colloidal suspensions \cite{Shi2013}. Acoustic Mie resonances have also been leveraged to design metamaterials operating at acoustic and ultrasonic frequencies \cite{Liu2000,Li2004,Fang2006,Brunet2013}. Mie scattering of phonons has been explored theoretically \cite{Prasher2004} but not applied to specific materials or to calculate thermal conductivity. There has been considerable effort to reduce $\kappa$ of semiconductors by randomly incorporating nanoparticles \cite{Zide2005,Kim2006a,Mingo2009}, but these studies do not explicitly consider Mie resonance. 

In this article, we evaluate the impact of acoustic Mie scattering of phonons on the thermal conductivity of crystalline semiconductor and amorphous polymer composites. First, we develop a model to evaluate the impact of acoustic Mie resonance on $\kappa$. Then we apply the model to ErAs nanoparticles in crystalline InGaAs alloy, a case in which the spectral thermal conductivity is known, and explain the previously measured reduction in $\kappa$. Our effort to extend the model to amorphous polymer composites yields inconclusive results, sensitive to the particular model used to calculate the spectral thermal conductivity of the amorphous polymer matrix.

The lattice or phonon contribution to $\kappa$ can be calculated using the Callaway model \cite{Callaway1959}, which is a solution to the Boltzman transport equation within the relaxation time approximation, assuming a linear phonon dispersion relation. The thermal conductivity is given by
\begin{equation}
\kappa=\frac{k_B}{2\pi^2v_1}\left(\frac{k_BT}{\hbar}\right)^3 \int_0^{\theta_D/T}\tau_{ph}(x) \frac{x^4e^x}{(e^x-1)^2}dx,
\end{equation}
where $k_B$ is Boltzman's constant, $\hbar$ is Planck's constant, $T$ is the temperature, $x$ is the normalized frequency $\hbar\omega/k_BT$, $v_1$ is the speed of longitudinal acoustic phonons, $\theta_D$ is the Debye temperature, and $\tau_{ph}$ is the relaxation time of phonons, which follows Matthiessen's rule
\begin{equation}
\tau_{ph}^{-1}=\tau_{n}^{-1}+\tau_{u}^{-1}+\tau_{a}^{-1}+\tau_{b}^{-1}+\tau_{e-ph}^{-1}+\tau_{p}^{-1},
\end{equation}
where $\tau_{n}$, $\tau_{u}$, $\tau_{a}$, $\tau_{b}$, $\tau_{e-ph}$, and $\tau_{p}$, are the relaxation times associated with normal phonon-phonon, umklapp, alloy, boundary, electron-phonon, and nanoparticle scattering, respectively. The relaxation time due to uncorrelated scattering by nanoparticles is given by \cite{Kim2006b,Kim2006a}
\begin{equation}
\tau_{p}^{-1}=v_1\sigma_p N,
\end{equation}
where $N$ is the volume density of nanoparticles and $\sigma_p$ is the scattering cross section. In the literature, $\sigma_p$ is approximated by an interpolating function connecting the Rayleigh cross section at low frequencies to the geometrical cross section at high frequencies \cite{Kim2006b,Mingo2009}. As a result, at intermediate frequencies the expression fails to capture the effects of resonant scattering due to the excitation of normal modes in the sphere.

To accurately capture acoustic resonances, it is necessary to use the full solution for the scattering cross section of acoustic waves by an elastic sphere \cite{Anderson1950,Morse1986},
\begin{equation}
\sigma_p=\frac{4\pi}{k_1^2}\sum_{m=0}^\infty(2m+1)\left| A_m \right|^2, \quad \text{where}
\end{equation}
$$
A_m=\frac{z_2j'_m(k_1a)j_m(k_2a)-z_1j'_m(k_2a)j_m(k_1a)}{z_2h'_m(k_1a)j_m(k_2a)-z_1j'_m(k_2a)h_m(k_1a)},
$$
$a$ is the sphere radius, $k_1=\omega/v_1$ and $k_2=\omega/v_2$ are the wavevectors in the host and sphere, respectively, $z_1=\rho_1v_1$ and $z_2=\rho_2v_2$ are the acoustic impedances, where $\rho_1$ and $\rho_2$ are the densities, and $h_m(\xi)=j_m(\xi)+iy_m(\xi)$, where $j_m(\xi)$ and $y_m(\xi)$ are the spherical Bessel functions, and the primes denote differentiation with respect to the argument. Acoustic resonances correspond to the poles of $A_m$. They are expected to occur when the wavelength in the background medium is much larger than the particle ($k_1a<<1$), ensuring that the phase of the incident field is constant across the particle. Setting the imaginary and real parts of the denominator of $A_m$ equal to zero, respectively, for $k_1a<<1$ and $m=0$ we obtain
\begin{equation}
\cos({k_2a})-\left(1-\frac{\rho_2}{\rho_1} \right)\frac{\sin(k_2a)}{k_2a}=0,
\end{equation}
\begin{equation}
\frac{\cos({k_2a})}{k_2a}-\frac{\sin(k_2a)}{(k_2a)^2}=0.
\end{equation}
For densities of $\rho_2/\rho_1=0$, $1$, and $\infty$, the lowest-frequency solutions of Eq. (5) are $k_2a=0$, $\pi/2$, and $\pi$, respectively, while the lowest-frequency solution of Eq. (6) is $k_2a\approx4.5$. Therefore, the fundamental resonant frequency depends on the particle density and is given by $k_2a\leq\pi$ or $\omega\leq\pi v_2/a$. As such, the fundamental resonant frequency increases as $\rho_2/\rho_1$ and $v_2$ increase, and decreases as the particle radius increases.

It is important to make a few observations. First, since Mie scattering is a resonance phenomenon, it will be effective only over a narrow bandwidth. Thus, to achieve large reductions in thermal conductivity, the shape of the spectral thermal conductivity, given by the integrand of Eq. (1), will play a vital role in determining which phonons to scatter. Second, our model implicitly assumes thermal transport is coherent, rather than diffusive. This, in turn, assumes the phonon mean free path at the frequency of interest is much longer than the particle size, a condition attainable in crystalline and inorganic compounds. However, for amorphous materials, additional information about the spectral thermal conductivity and the frequency-dependent mean free path will be required to establish the validity of our formalism.

\begin{figure}
\includegraphics[scale=0.6]{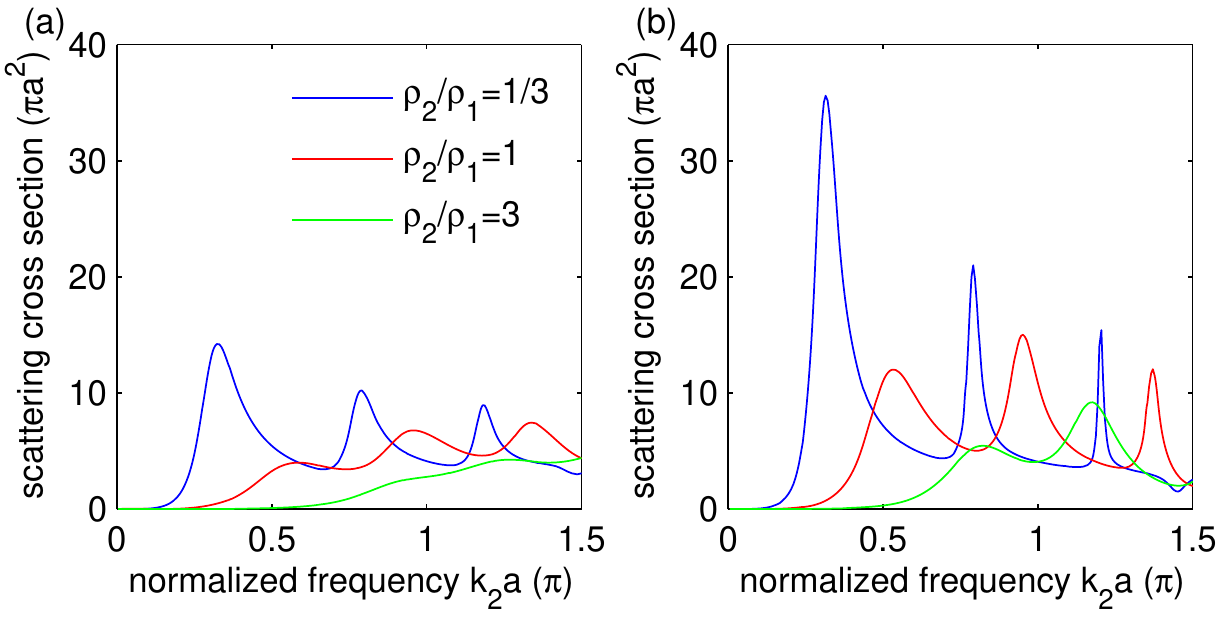}
\caption{\label{fig:epsart} Scattering cross section versus normalized frequency $k_2a$ for spheres with densities of $\rho_2/\rho_1=$ 1/3, 1, and 3 and sound speeds of (a) $v_2/v_1=1/2$ and (b) $v_2/v_1=1/3$.}
\end{figure}

To further explore the dependence of the scattering cross section on the acoustic properties of the particles, Eq. (4) was used to calculate $\sigma_p$ versus the normalized frequency $k_2a$ for spheres with densities of $\rho_2/\rho_1=1/3$, 1, and 3 and sound speeds of (a) $v_2/v_1=1/2$ and (b) $v_2/v_1=1/3$ (Fig. 1). Consistent with the discussion following Eq. (5), the fundamental resonant frequency, seen as the lowest-frequency peak in Fig. 1, decreases as the sound velocity and density of the sphere decrease. The calculations also indicate that the magnitude of the cross section at the acoustic resonances increases as $v_2/v_1$ and $\rho_2/\rho_1$ decrease, suggesting that a large mismatch of the acoustic properties is necessary to obtain large scattering cross sections. Moreover, since the cross section at the resonant frequencies greatly exceeds the geometrical cross sectional area of the sphere, the scattering rate $\tau_{p}^{-1}$, given by the product $\sigma_p N$, can be large even when the density of nanoparticles is small. This is an important advantage for thermoelectric materials because it allows the thermal conductivity to be reduced with minimal impact on the electrical conductivity.

To understand the physical origin of the scattering resonances, Fig. 2 shows the magnitude of the total pressure field around a particle with $\rho_2/\rho_1=v_2/v_1=1/3$ at frequencies corresponding to the first two peaks of the scattering cross section in Fig 1(b). The calculations assume that both the pressure field and the normal component of the particle velocity, proportional to the pressure gradient, are continuous at the boundary of the sphere \cite{Anderson1950}. At the fundamental resonance frequency ($k_2a/\pi=0.3$), shown in Fig. 2(a), the scattered pressure field is spherically symmetric and corresponds to an acoustic monopole. The resonance at $k_2a/\pi=0.8$ shows two adjacent regions of high pressure [Fig. 2(b)], corresponding to an acoustic dipole.

\begin{figure}
\includegraphics[scale=0.7]{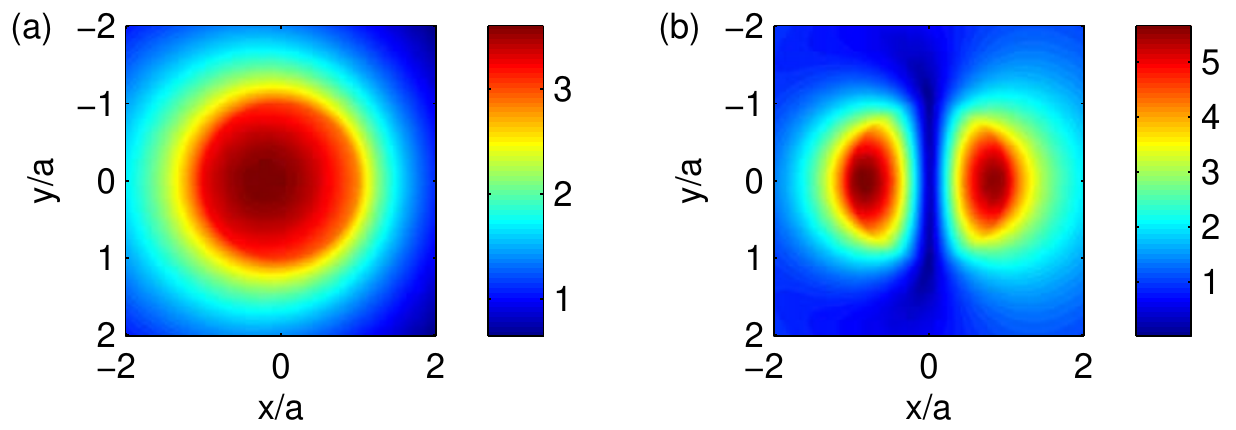}
\caption{\label{fig:epsart} Magnitude of the total pressure field around a particle with $\rho_2/\rho_1=v_2/v_1=1/3$ at the (a) acoustic monopole ($k_2a/\pi=0.3$) and (b) acoustic dipole ($k_2a/\pi=0.8$).}
\end{figure}

The results of Fig. 1 indicate that large acoustic scattering cross sections are possible when the sound speed and mass density of the particle are much smaller than those of the host material. This implies that the acoustic impedance and bulk modulus of the particle, given by $\rho_2v_2$ and $ \rho_2 v_2^2$, respectively, should be minimized to enhance scattering. Since semiconductor compounds and alloys have similar sound speeds and densities, semiconductor nanoparticles are not expected to produce strong scattering resonances when embedded in a semiconductor host. On the other hand, soft ductile materials such as rubidium (Rb) have extraordinarily low bulk modulus, and should in principle exhibit acoustic resonances when embedded in a semiconductor.

\begin{table}
\caption{Parameters used in the calculations.}
\begin{ruledtabular}
\begin{tabular}{l c c c}
Material & Density (kg/m$^3$)& Sound speed (m/s) & Ref.\\
\hline
In$_{0.53}$Ga$_{0.47}$As &  5506  & 4253 & \cite{Goldberg1999}  \\
ErAs &  8550  & 2460 & \cite{Singh2016}\\
Rb  &  1580 & 1370 &\cite{Gutman1967} \\
Polysiloxane  &  1300 & 1070 &\cite{Still2013} \\
Polybutylene  &  930 & 1690 &\cite{Freeman1986} \\
\end{tabular}
\end{ruledtabular}
\end{table}

\begin{figure}
\includegraphics[scale=0.68]{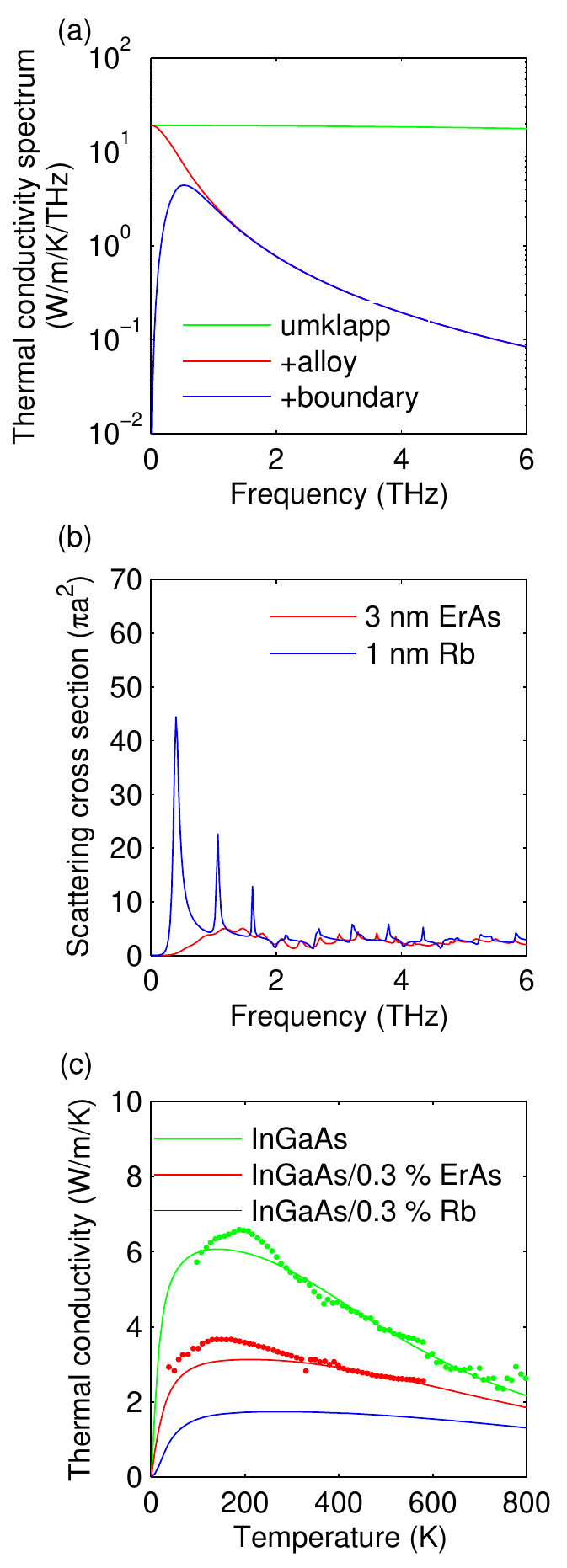}
\caption{\label{fig:epsart} (a) Spectral thermal conductivity of InGaAs at \\300 K due to the predominant phonon scattering processes, (b) scattering cross section versus frequency for 2.4 nm ErAs and 1 nm Rb nanoparticles in InGaAs, and (c) the corresponding temperature dependence of thermal conductivity for a volume fraction of 0.3 \%. Data are from Ref. \cite{Kim2006a}.}
\end{figure}

To demonstrate the role of acoustic resonances in reducing thermal conductivity, we performed calculations for Rb particles in In$_{0.53}$Ga$_{0.47}$As using the material properties in Table 1. Figure 3(a) shows the thermal conductivity spectrum of InGaAs alloy, given by the integrand of Eq. (1). The relaxation times used in the calculations, $\tau_{u}^{-1}=2\times10^{-24}\omega^2 T^3\exp{(-\theta_D/3T)}$, $\tau_{a}^{-1}=8\times10^{-42}\omega^4$, and $\tau_{b}^{-1}=10^{9}$ s$^{-1}$, were chosen to fit the temperature-dependent thermal conductivity measured in Ref. \cite{Kim2006a}. When only umklapp processes are present, the thermal conductivity spectrum is flat (green), indicating that all phonons with frequencies less than the Debye frequency (=6.9 THz) contribute to thermal conduction. When alloy scattering is also included, high-frequency phonons are scattered and the thermal conductivity spectrum shifts to lower frequencies (red). When all mechanisms are present, including boundary scattering, the remaining spectrum consists of phonons with frequencies ranging from 0.5-1.5 THz (blue) with a maximum around 0.5 THz.

The remaining portion of the thermal conductivity spectrum can be reduced by introducing nanoparticles. For example, Fig. 3(b) shows the scattering cross section for 3-nm ErAs particles in InGaAs (red), representing the state of the art in nanoparticle scattering \cite{Kim2006a}. For this case, $\sigma_p$ is low and concentrated around 2 THz where the thermal conductivity spectrum is small [Fig. 3(a)]. On the other hand, calculations for 1-nm Rb nanoparticles, which have approximately 10 times smaller bulk modulus than ErAs, produce scattering efficiencies greater than $50\pi a^2$. Note this effect is strongly material dependent and cannot be obtained by simply changing the size of ErAs nanoparticles.

To calculate the corresponding reduction in thermal conductivity, the scattering cross sections in Fig. 3(b) were used to calculate the relaxation time from Eq. (3). Figure 3(c) shows the calculated $\kappa$ for InGaAs without nanoparticles (green), InGaAs with 0.3\% 3 nm ErAs particles (red), and InGaAs with 0.3\% 1-nm Rb particles (blue).  The magnitude and temperature dependence of thermal conductivity with and without ErAs are in excellent agreement with published data (dots) \cite{Kim2006a}. With ErAs particles, the thermal conductivity at room temperature is reduced by a factor of 1.4, from 5.5 W/m/K to 4 W/m/K. For the same loading of 0.3\%, 1-nm Rb nanoparticles reduce the $\kappa$ to 1.75 W/m/K, greater than a factor of three reduction compared to the alloy. Such a steep reduction in thermal conductivity is due to the large magnitude of the scattering cross section and the alignment of the resonance with the peak of the thermal conductivity spectrum. We also find that if the loading is increased further to 3\%, the thermal conductivity reduces to 0.88 W/m/K.

Amorphous polymers have the lowest thermal conductivity of the solid materials. Reducing their thermal conductivity further would greatly benefit a number of insulation applications. However, our model must be applied with caution in this case because the spectral thermal conductivity of amorphous materials is not well understood or available. Picosecond ultrasound spectroscopy provides evidence that long-wavelength propagating acoustic phonons are present in amorphous polymers \cite{Morath1996,Pontecorvo2011}. These measurements indicate a $1/\omega^2$ dependence of the mean free path up to 320 GHz. At higher frequencies, Rayleigh scattering by disorder it is expected to dominate as it does in amorphous SiO$_2$ \cite{Vacher1997,Masciovecchio2006}. 

\begin{figure}
\includegraphics[scale=0.65]{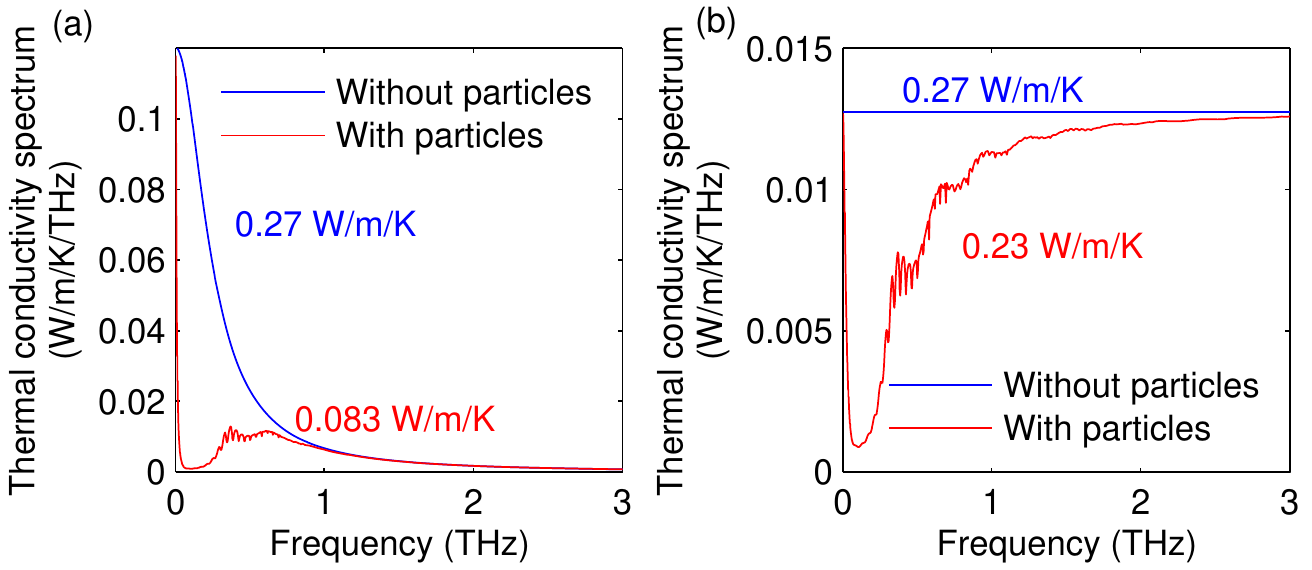}
\caption{\label{fig:epsart} Calculated thermal conductivity spectrum for amorphous polymers with (red) and without (blue) 10\% loading of 10-nm polysiloxane nanoparticles, using (a) using theoretical expressions for scattering rates and (b) flat phonon spectrum.}
\end{figure}

Since the spectral thermal conductivity of amorphous polymers is unknown beyond 320 GHz, we consider two cases. First, shown by the blue line in Fig. 4(a), we calculate the spectral thermal conductivity for polybutylene using theoretical expressions for the $1/\omega^2$ \cite{Woodruff1961} and Rayleigh scattering rates \cite{Zaitlin1975}. In this case, the thermal conductivity spectrum is concentrated at frequencies less than 1 THz, where nanoparticle scattering can be effective. The calculated $\kappa$ is 0.27 W/m/K. In the second case, we assume the observed $1/\omega^2$ dependence of the mean free path holds up to the Debye frequency, and adjust the scattering rate to obtain the same thermal conductivity. This represents the worst-case scenario of a constant thermal conductivity spectrum, shown by the blue line in Fig. 4(b). 

We now consider adding nanoparticles. For the nanoparticle, we use polysiloxane, a common choice in acoustic metamaterials due to its very low sound speed (1070 m/s) \cite{Still2013}. The red line in Fig. 4(a) shows the modification of the thermal spectrum by 10\% loading of 10-nm polysiloxane particles. In this case, the effect is significant, leading to a near 70\% reduction to 0.083 W/m/K. The reduction is large in this case because the thermal conductivity spectrum is concentrated where the nanoparticle scattering is large. When the particles are added to the matrix with a flat spectrum, shown by the red line in Fig. 4(b), only phonons with frequencies less than 1 THz are removed, leaving most of the spectrum unaffected. As a result, the thermal conductivity is only reduced by 15\%, to 0.23 W/m/K. These results emphasize the importance of the thermal spectrum in determining the impact of acoustic resonances on thermal conductivity of amorphous polymer composites. An experimental determination of the thermal spectrum of amorphous polymers will be required to obtain a more reliable design with ultralow thermal conductivity.

Nanoparticles are a proven method to scatter phonons and reduce thermal conductivity. However, to date the research has focused primarily on the lattice match between the particle and host, while the dependence of the phonon scattering on the acoustic properties and size of the nanoparticles remains largely unexplored. Moreover, due to the approximations in current models of nanoparticle scattering, the effects of acoustic resonance have not been investigated. This article explores the impact of acoustic resonances in nanoparticles on the phonon scattering and thermal conductivity. Phenomenological models based on continuum acoustic theory indicate that by appropriately choosing the size and acoustic properties of the particles, namely the mass density and bulk modulus, acoustic monopole and dipole resonances in spherical nanoparticles can lead to greatly enhanced scattering cross sections, providing a large reduction in thermal conductivity with a small concentration of nanoparticles, limiting the negative impact on electrical conductivity and maximizing $ZT$ for thermoelectric applications. Our efforts to apply this model to amorphous polymers yielded inconclusive results due to the uncertainty in the spectral thermal conductivity of amorphous polymers. Systematic measurements of the spectral thermal conductivity of amorphous polymers will enable a more reliable design with reduced thermal conductivity.

This work was supported by ARPA-E under Contract No. DE-AR0000735.

\bibliography{bib}
\end{document}